# SMISS: A protein function prediction server by integrating multiple sources

Renzhi Cao[1], Zhaolong Zhong [1], and Jianlin Cheng[1, 2, 3, *]

[1]Department of Computer Science, University of Missouri, Columbia, Missouri, 65211, USA. [2]Informatics Institute, University of Missouri, Columbia, Missouri, 65211, USA. [3]C. Bond Life Science Center, University of Missouri, Columbia, Missouri, 65211, USA.

*To whom correspondence should be addressed.
Email: chengji@missouri.edu

**Abstract:** SMISS is a novel web server for protein function prediction. Three different predictors can be selected for different usage. It integrates different sources to improve the protein function prediction accuracy, including the query protein sequence, protein-protein interaction network, gene-gene interaction network, and the rules mined from protein function associations. SMISS automatically switch to ab initio protein function prediction based on the query sequence when there is no homologs in the database. It takes fasta format sequences as input, and several sequences can submit together without influencing the computation speed too much. PHP and Perl are two primary programming language used in the server. The CodeIgniter MVC PHP web framework and Bootstrap front-end framework are used for building the server. It can be used in different platforms in standard web browser, such as Windows, Mac OS X, Linux, and iOS. No plugins are needed for our website. Availability: http://tulip.rnet.missouri.edu/profunc/.

# Introduction

Protein structure and function prediction has a lot of biomedical and pharmaceutical implications (Radivojac et al. 2013; Cao et al. 2014a; Cao et al. 2014b; Cao et al. 2015a; Cao et al. 2015b; Li et al. 2015; Cao and Cheng 2016a), and it helps people understand their life. Most biologist use tool Basic Local Alignment Search Tool (BLAST) to search the query sequence against the PDB database, and the functions of homologs are used for the protein function prediction (Wang et al. 2013; Cao and Cheng 2016a). Several methods apply it for protein function prediction, such as GOtcha (Jo and Cheng 2014), OntoBlast (Jo 2015), and Goblet (Schölkopf and Smola 2002). But it may fail when there is few or no homologs found in the database. Kihara and his groups developed a webserver PFP (Hawkins et al. 2009) which uses profile-sequence alignment tool PSI-BLAST (Rost 1998) to get more sensitive predictions. However, there are several other sources to infer more protein functions, such as protein-protein interaction network (Yang and Zhou 2008a; Zhang and Zhang 2010; Wang et al. 2011; Li et al. 2014; Adhikari et al. 2015; Cao and Cheng 2016b), gene-gene interaction network (Chen et al. 2010; Cao and Cheng 2015), protein structure data (Cao et al. 2015c; Deng et al. 2015; Kryshtafovych et al. 2015), microarrays (Uziela and Wallner 2016), and combination of different sources (Wu et al. 2007; Yang and Zhou 2008b; Zhou and Skolnick 2011; Eickholt and Cheng 2012), etc. Also, there are some general websites for protein structure prediction which also includes the protein function prediction, such as I-TASSER (Yang et al. 2015). But since this is not for protein function prediction specific, it takes longer time to make prediction. Our SMISS server uses different sources and combines them for protein function prediction, including protein sequences, function associations, and protein-protein or gene-gene interaction networks (Cao and Cheng 2015; Cao and Cheng 2016a). It can take use of homologs function, and also the protein sequence itself for ab initio protein function prediction. More importantly, the user can choose different sources for the function prediction based on their need.

## Implementation and usage of the website

SMISS webserver has been implemented in PHP and perl. **Figure 1** shows the home page of the webserver. The front end language is PHP, and the background programs for calculating the protein function is done by perl. The program background code can be requested from the author (Cao and Cheng 2016a). CodeIgniter MVC web framework is used for our website. It is a simple and elegant toolkit for creating full-featured applications. MVC controller system is used in our website. Users can register in our website. **Figure 2** shows the registration webpage. After registration, users can login the website, and the website saves all results for them in case they need it in the near future.

In order to save the computation cost, at most 5 jobs per day can be submitted for non-registered user. The registered user can submit more jobs every day. The server will suspend some jobs later submit when the running jobs is overloaded for the server. We also apply Bootstrap front-end framework to this server to provide efficient and neat user interface.

We put the job submission as the main feature in the home page. There are two ways to input the protein sequence. First, you can copy one or multiple sequences in fasta format to the “Paste a fasta” text box. Second, you can upload a protein sequence file in fasta format from your local machine.

After that, you can fill in your email address, and choose a method for prediction. There are three options for the users to choose. The first one is SMISS, which includes all sources for protein function prediction. The second one is MIS score, which only uses the source from the hits by the tool PSI-BLAST. The third one is MIS-NET score, which uses the hits by the tool PSI-BLAST and the network sources (Cao and Cheng 2016a). The user can choose the suitable one for their requirement.

## Job processing

After user submits a job, the webpage will be redirected to job processing page which displays the status of submitted job. The job processing page is shown in **Figure 3**. The job status is updating every five seconds, and result will be shown in this webpage after the job finishes. Users can also close this webpage and get the results by using the last URL or link. **Figure 4** shows the job queue of all submitted jobs. The user can check the status of their job in this website. The jobs status can be Pending, Running, and Finished. Some jobs maybe pending because too many jobs already submitted to the server. In addition, users can also check the user's map showing the location of all visitors in the world. It is shown in **Figure 5**. Moreover, users can also check the job history of all jobs submitted by themselves, which is shown in **Figure 6**.

## Function prediction result

**Figure 7** shows the prediction results when the job finishes. The left column shows the top 10 function prediction GO terms based on the confidence score. The right columns shows the percentage of the function prediction in each GO terms category (BP-biological process, MF-molecular function, CC-cellular component). You can get all prediction results by clicking "Check all results" button.

## Conclusions

In this paper, we introduce a novel protein function prediction website SMISS. We integrate information from different sources for protein function prediction, such as the one from profile-sequence search tool PSI-BLAST, from the knowledge learned of the true function by data mining techniques, from gene-gene/protein-protein interaction networks, and from the protein sequence itself. Our web server can make ab-initio function predictions from the query sequence, even there is no homologs found by traditional sequence-profile search tool PSI-BLAST. Users are free choose which combination of sources for the protein function prediction based on their need. The speed is not influenced too much when you submit several fasta in one file, which is good for the users who needs to make protein function prediction for large data. In the future, we plan to improve the user interface to make it more friendly, and also make the website more robust for the security issue. In addition, we want to improve the speed and accuracy of function prediction method for other people to better use our website.

## Competing interests

The authors declare that they have no competing interests.

# Figures

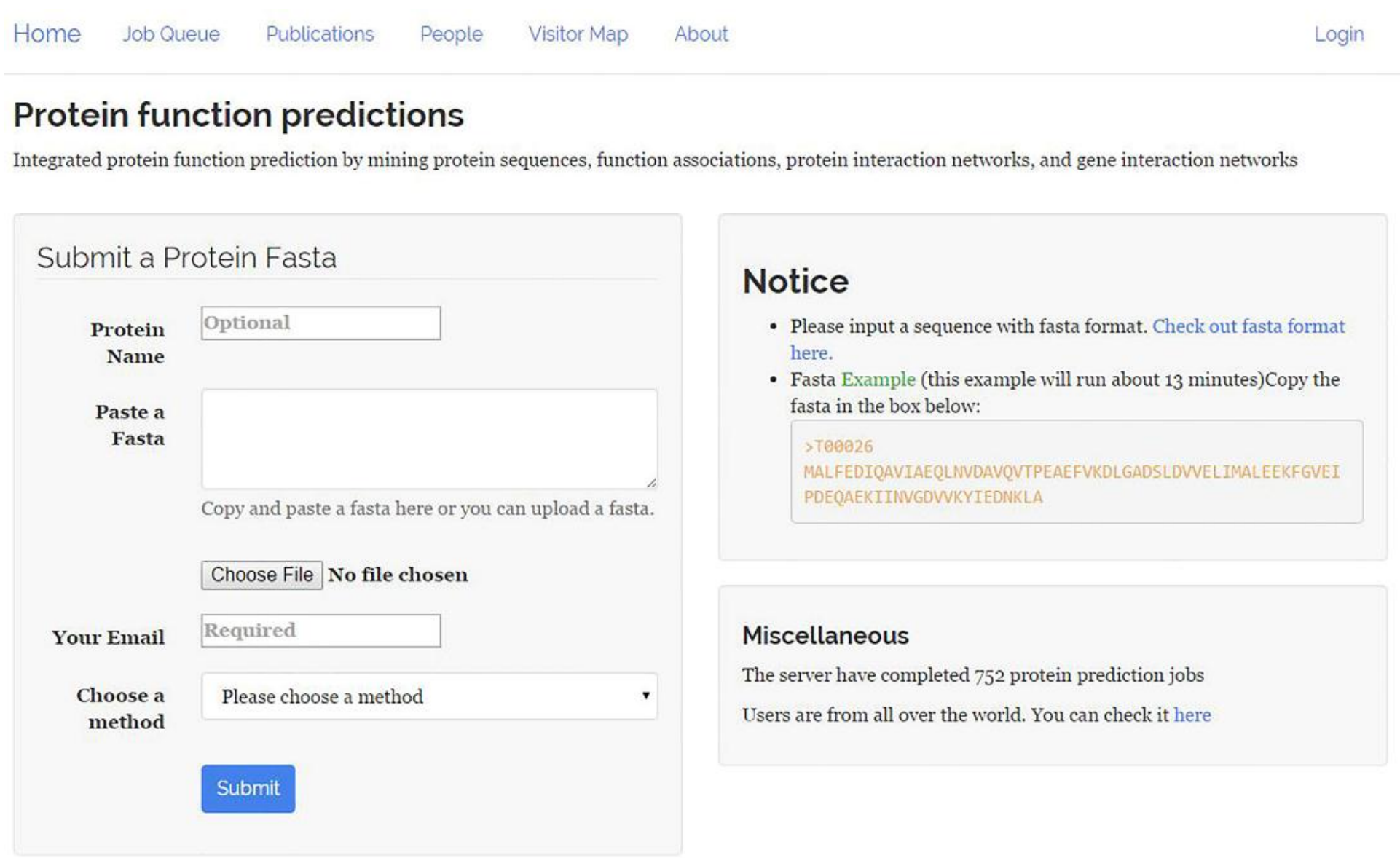

Figure 1. The home page of SMISS webserver.

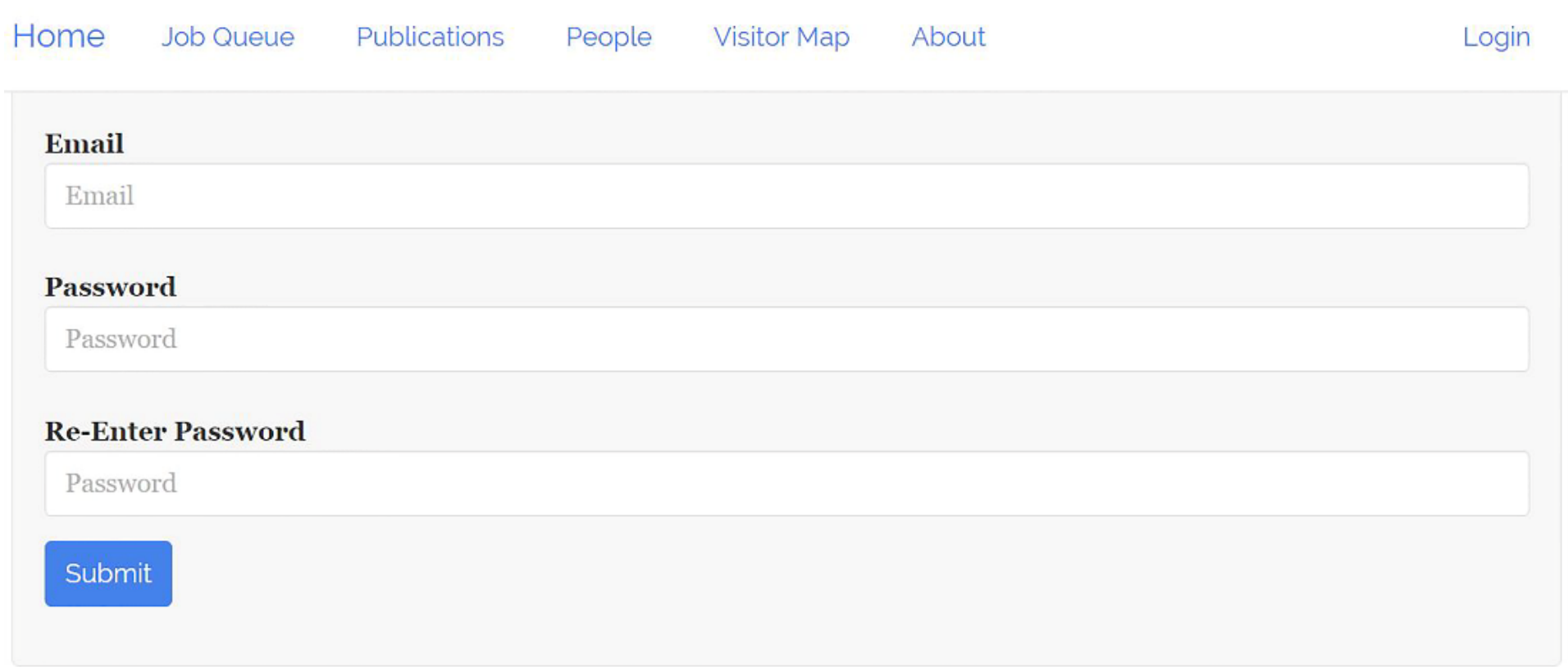

Figure 2. The registration page of SMISS webserver.

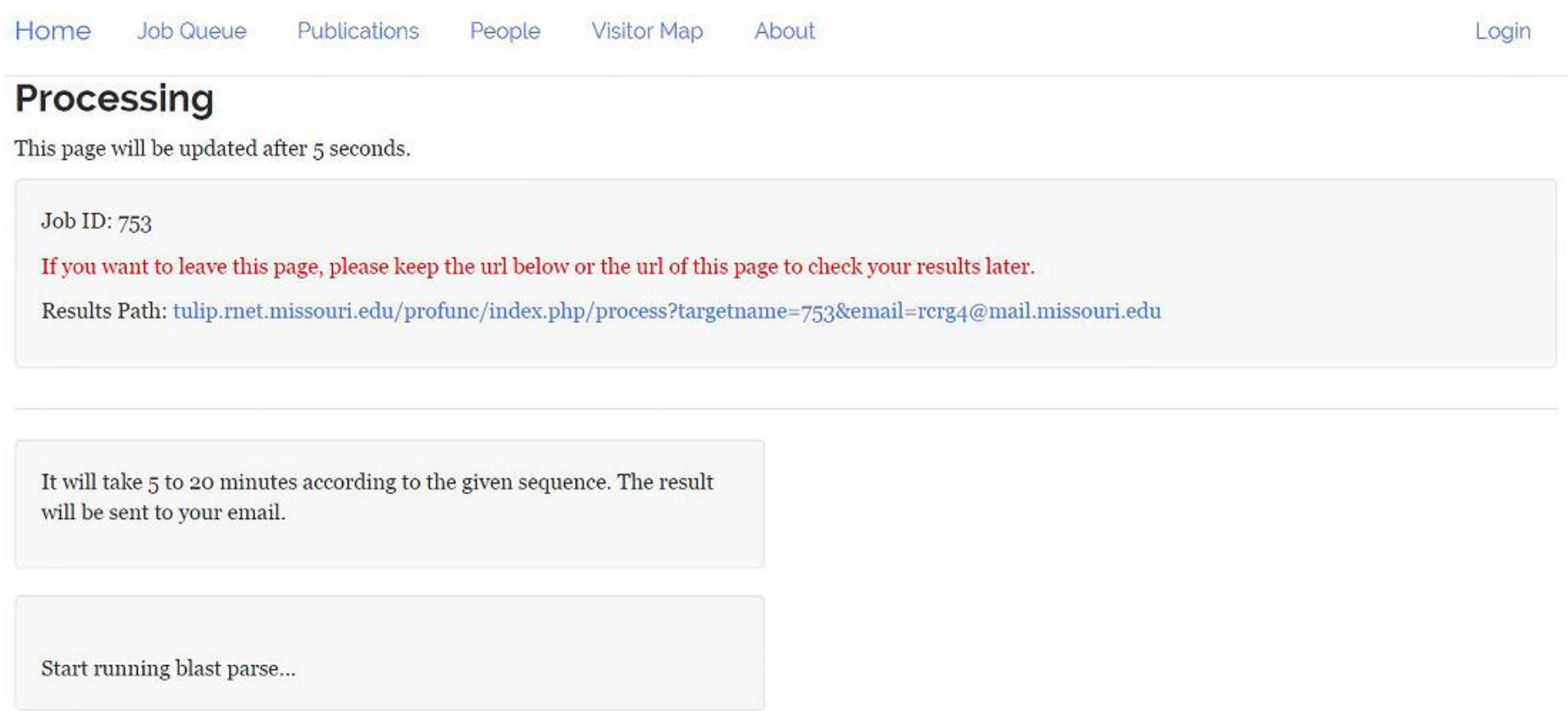

Figure 3. The job processing page of SMISS webserver.

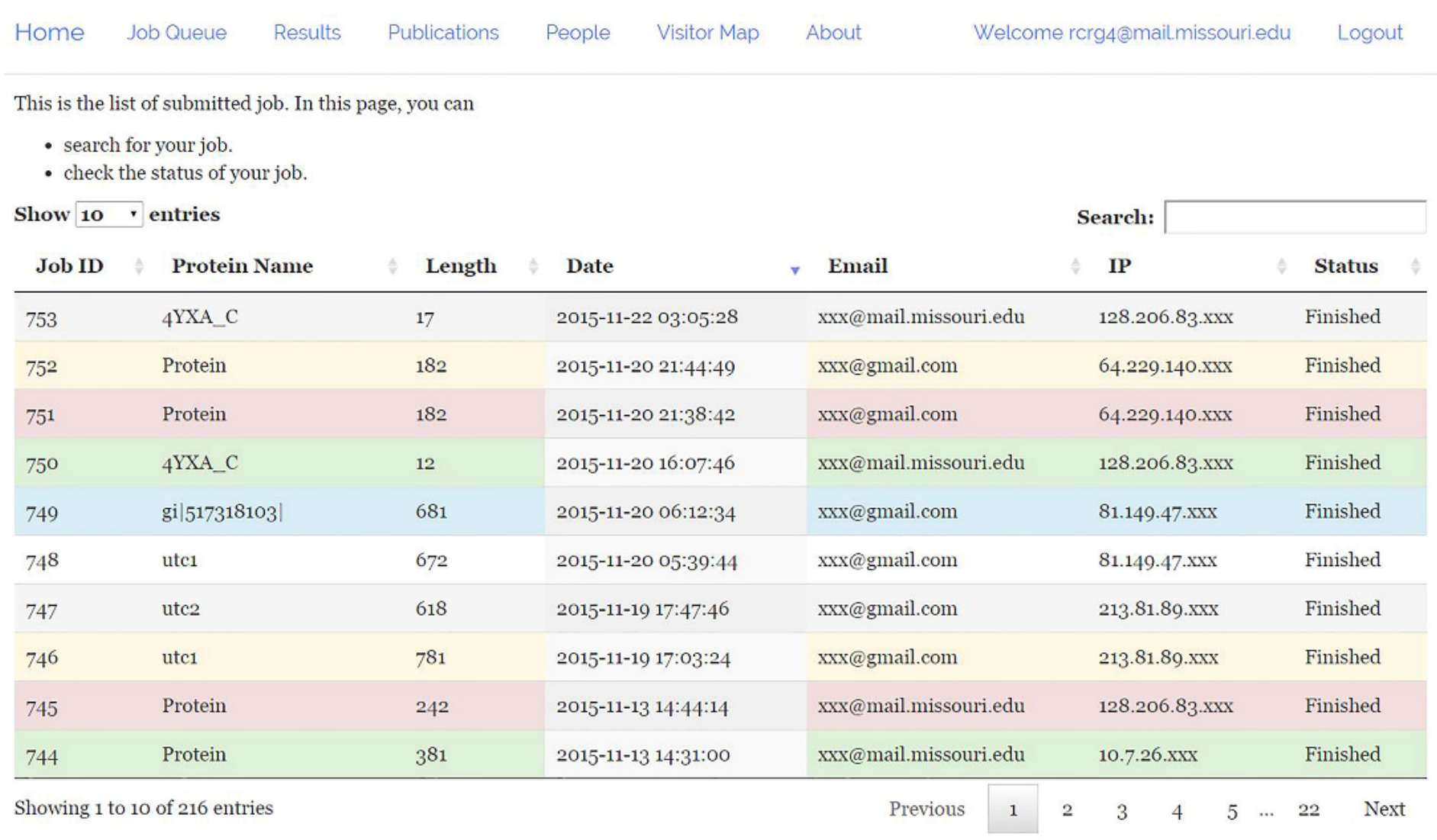

Home Job Queue Results Publications People Visitor Map About Welcome rcrg4@mail.missouri.edu Logout

This is the list of submitted job. In this page, you can

- search for your job.
- check the status of your job.

**Show** 10 **entries** **Search:**

| Job ID | Protein Name | Length | Date | Email | IP | Status |
|---|---|---|---|---|---|---|
| 753 | 4YXA_C | 17 | 2015-11-22 03:05:28 | xxx@mail.missouri.edu | 128.206.83.xxx | Finished |
| 752 | Protein | 182 | 2015-11-20 21:44:49 | xxx@gmail.com | 64.229.140.xxx | Finished |
| 751 | Protein | 182 | 2015-11-20 21:38:42 | xxx@gmail.com | 64.229.140.xxx | Finished |
| 750 | 4YXA_C | 12 | 2015-11-20 16:07:46 | xxx@mail.missouri.edu | 128.206.83.xxx | Finished |
| 749 | gi\|517318103\| | 681 | 2015-11-20 06:12:34 | xxx@gmail.com | 81.149.47.xxx | Finished |
| 748 | utc1 | 672 | 2015-11-20 05:39:44 | xxx@gmail.com | 81.149.47.xxx | Finished |
| 747 | utc2 | 618 | 2015-11-19 17:47:46 | xxx@gmail.com | 213.81.89.xxx | Finished |
| 746 | utc1 | 781 | 2015-11-19 17:03:24 | xxx@gmail.com | 213.81.89.xxx | Finished |
| 745 | Protein | 242 | 2015-11-13 14:44:14 | xxx@mail.missouri.edu | 128.206.83.xxx | Finished |
| 744 | Protein | 381 | 2015-11-13 14:31:00 | xxx@mail.missouri.edu | 10.7.26.xxx | Finished |

Showing 1 to 10 of 216 entries Previous 1 2 3 4 5 ... 22 Next

Figure 4. The job queue page of SMISS webserver.

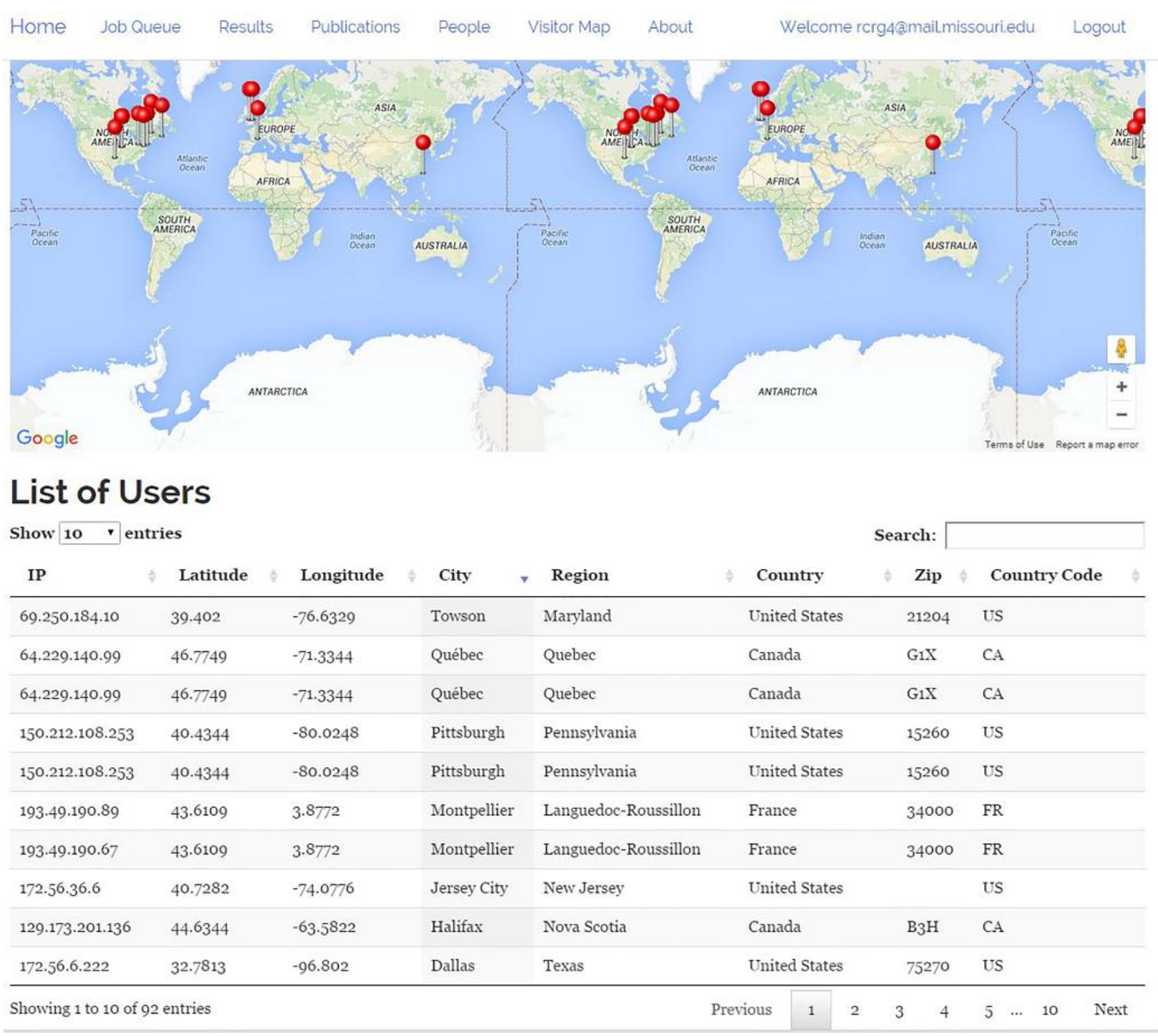

| IP | Latitude | Longitude | City | Region | Country | Zip | Country Code |
|---|---|---|---|---|---|---|---|
| 69.250.184.10 | 39.402 | -76.6329 | Towson | Maryland | United States | 21204 | US |
| 64.229.140.99 | 46.7749 | -71.3344 | Québec | Quebec | Canada | G1X | CA |
| 64.229.140.99 | 46.7749 | -71.3344 | Québec | Quebec | Canada | G1X | CA |
| 150.212.108.253 | 40.4344 | -80.0248 | Pittsburgh | Pennsylvania | United States | 15260 | US |
| 150.212.108.253 | 40.4344 | -80.0248 | Pittsburgh | Pennsylvania | United States | 15260 | US |
| 193.49.190.89 | 43.6109 | 3.8772 | Montpellier | Languedoc-Roussillon | France | 34000 | FR |
| 193.49.190.67 | 43.6109 | 3.8772 | Montpellier | Languedoc-Roussillon | France | 34000 | FR |
| 172.56.36.6 | 40.7282 | -74.0776 | Jersey City | New Jersey | United States | | US |
| 129.173.201.136 | 44.6344 | -63.5822 | Halifax | Nova Scotia | Canada | B3H | CA |
| 172.56.6.222 | 32.7813 | -96.802 | Dallas | Texas | United States | 75270 | US |

Figure 5. The location page of SMISS webserver.

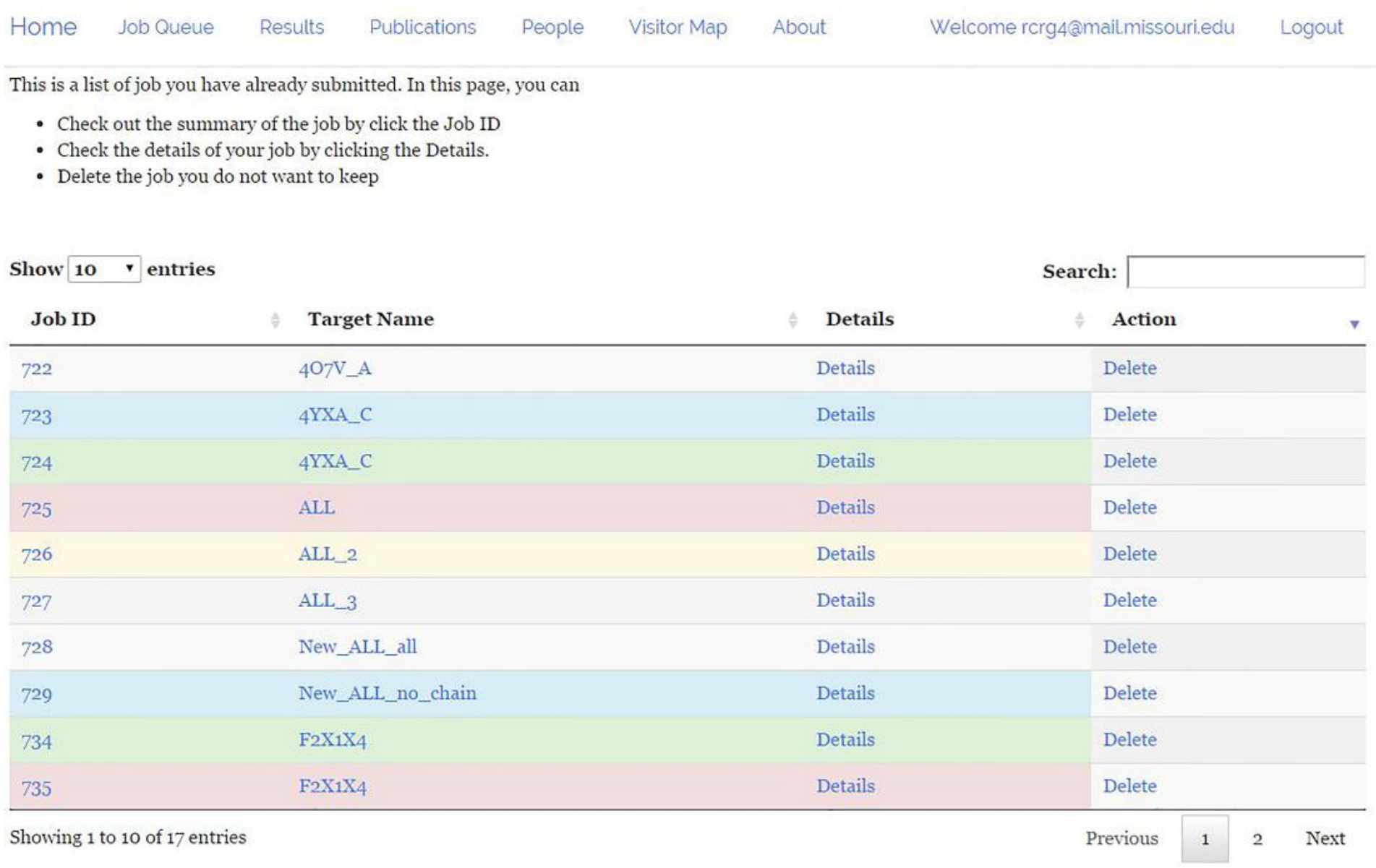

Figure 6. The job history page of SMISS webserver.

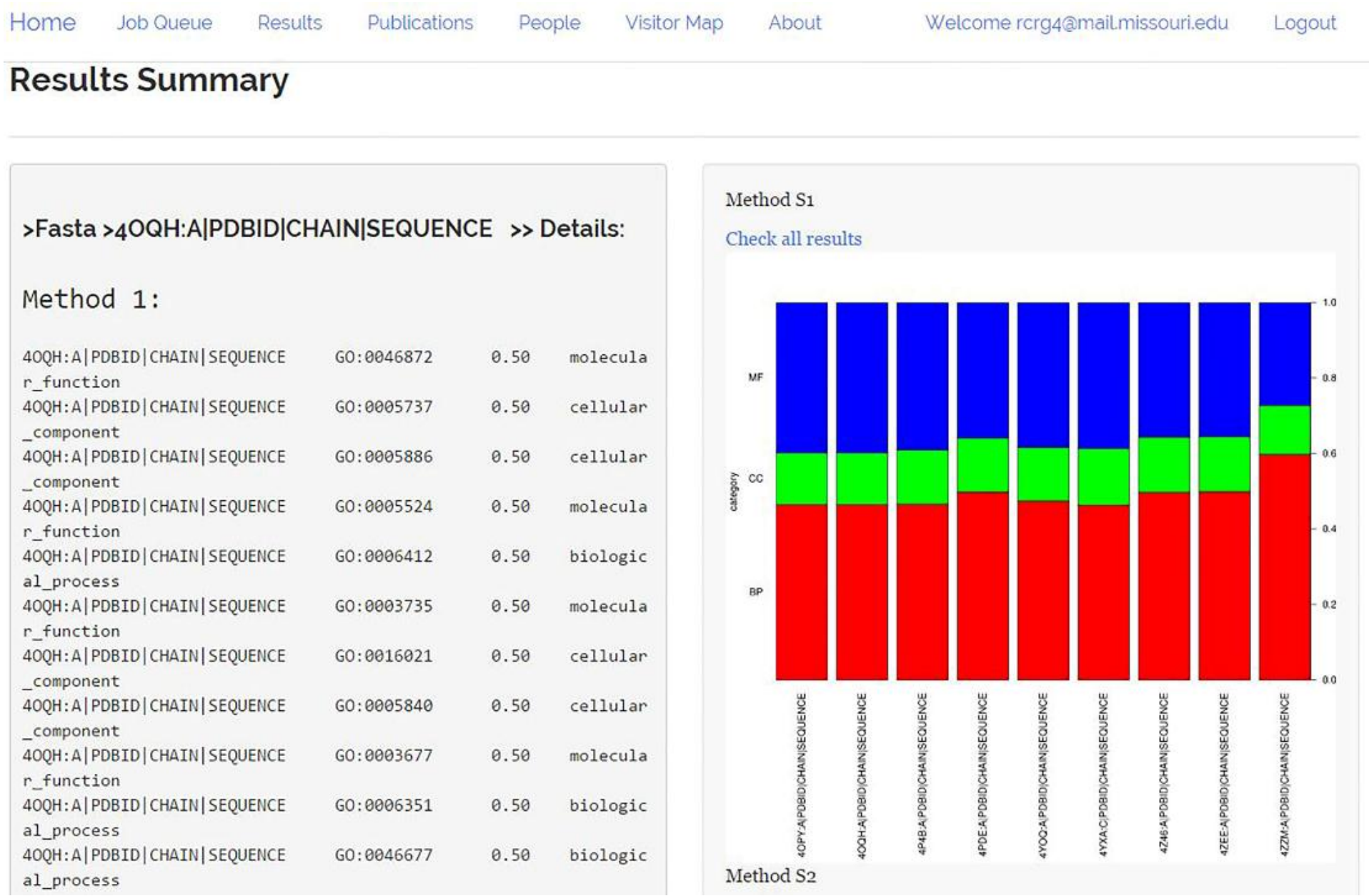

Figure 7. The prediction result page of SMISS webserver.